\documentclass[aps,prl,twocolumn,superscriptaddress,longbibliography]{revtex4-2}
\usepackage{amsmath,amssymb}
\usepackage[pdftex]{hyperref,graphicx}
\hypersetup{colorlinks = true, urlcolor = blue, linkcolor = blue, citecolor = blue}
\usepackage{physics}
\usepackage{xcolor}
\usepackage{bm}
\usepackage{ulem}

\begin{document}
\title{Disorder effects on the so-called Andreev band in Majorana nanowires}
\begin{abstract}
    We comment on a recent publication Phys. Rev. Lett. 130, 207001 (2023), pointing out that the periodic model for the superconducting gap and/or the spin splitting used by the authors is artificial and does not apply to any real systems.  In addition, we show that the resulting Andreev band introduced by this artificial and unrealistic periodicity is suppressed by the potential disorder invariably present in all experimental systems.  The results of this model are therefore contrived and do not apply to any experimental system.
\end{abstract}


\author{Sankar Das Sarma}
\affiliation{Condensed Matter Theory Center and Joint Quantum Institute, Department of Physics, University of Maryland, College Park, Maryland 20742, USA}
\author{Haining Pan}
\affiliation{Department of Physics, Cornell University, Ithaca, NY 14850, USA}
\maketitle

\begin{figure}[ht]
\centering
\includegraphics[width=3.4in]{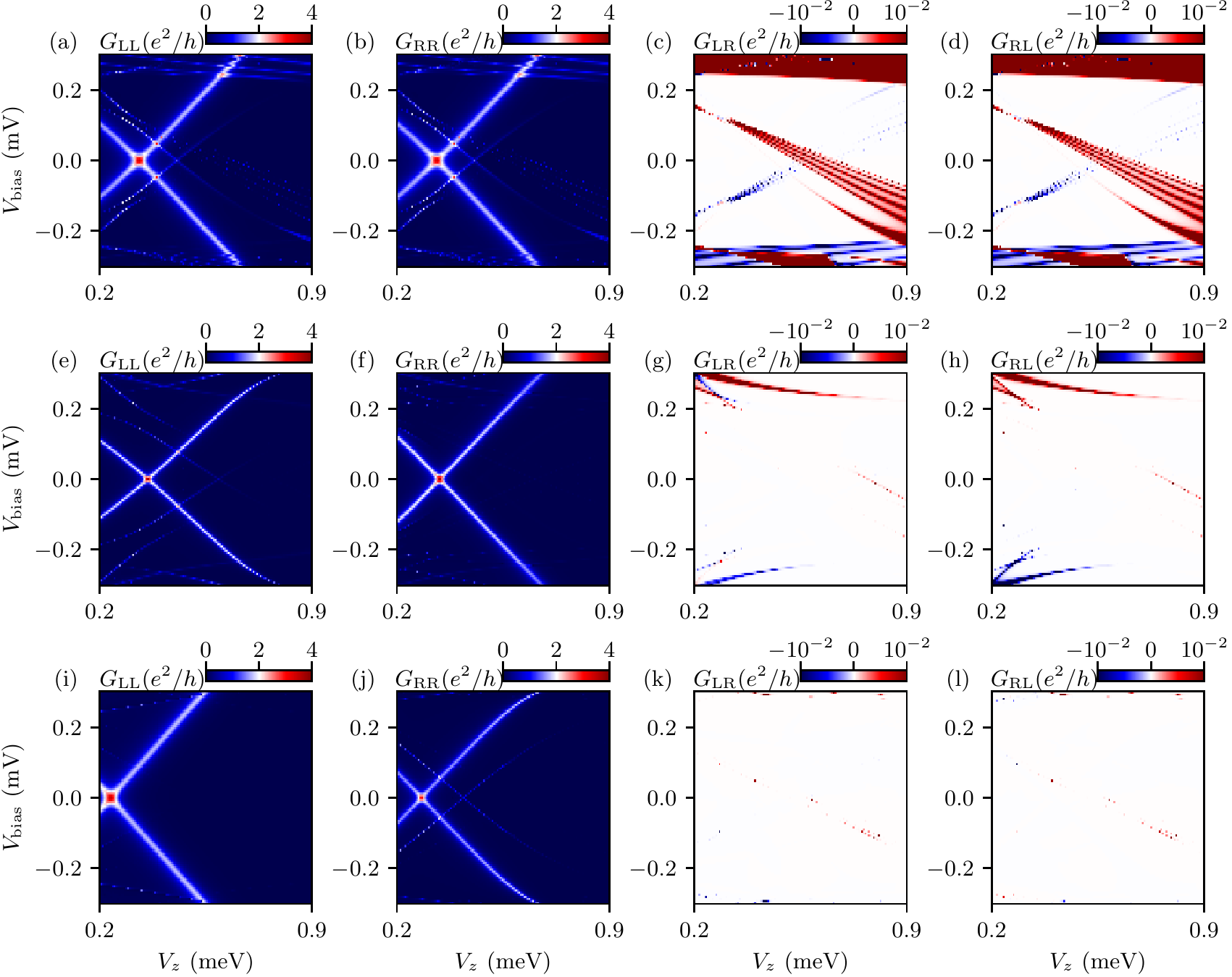}
\caption{
    Left (right) two panels show calculated local (nonlocal) conductance from both ends in wires with bulk trivial Andreev bands in the pristine system (as in  ~\cite{hess2023trivial}) by a finetuned periodic array of varying $g$-factor.  
    (a-d) pristine results showing physics in Ref.~\onlinecite{hess2023trivial} with ZBCPs in local tunneling from both ends and gap closing/opening-like features in nonlocal conductance.
    (e-h)/(i-l): results with a small (larger) chemical potential disorder of 0.5 (1) meV.
    Features in~\cite{hess2023trivial} are suppressed in (k-l).  
    All features are nontopological, and ZBCPs in local tunneling stem from manually inserted quantum dots at wire ends (Ref.~\onlinecite{hess2023trivial}). 
    The 3-$\mu$m wire has a bulk chemical potential of 1 meV and 0.5 meV at both ends (dot length scale $0.3~\mu$m) with random chemical potential disorder. 
    The $g$-factor profile comprises 5 periodic peaks of $g^*=1$, dividing the wire into disjointed parts spaced by $0.4~\mu$m, and zero elsewhere except ends set to 1.
    The spin-orbit coupling is 0, and the induced superconductivity is 0.2 meV. 
}
\label{fig:1}
\end{figure}

We point out that the effects of so-called Andreev band physics in Majorana nanowires proposed in~\cite{hess2023trivial} are suppressed by realistic disorder.  
Reference~\onlinecite{hess2023trivial} introduces an Andreev band into the bulk nanowire by putting in an artificial periodic (Fig. 1 in~\cite{hess2023trivial}) array of superconducting gap and/or Zeeman splitting throughout the nanowire and by assuming vanishing spin-orbit coupling, causing a seeming closing/opening of a gap-like structure in the nonlocal tunneling conductance measured across the wire ends. 
Additionally, Ref.~\onlinecite{hess2023trivial} also manually introduces another feature in the local tunneling, leading to a zero-bias conductance peak (ZBCP) in the tunneling spectroscopy. 
These two features appear incompatible with the methodology of using local and nonlocal tunneling spectroscopies together~\cite{rosdahl2018andreev,pan2021threeterminal} to ascertain topological Majorana zero modes in nanowires.
It is well-established that Majorana nanowires are dominated by random unintentional disorder~\cite{pan2020physical,dassarma2021disorderinduced,ahn2021estimating}, and the artificial finetuned Andreev band results presented in~\cite{hess2023trivial} are suppressed by disorder. We show our calculated representative results in Fig.~\ref{fig:1}, depicting Andreev band results for a periodic array of varying $g$-factor in the bulk nanowire (following~\cite{hess2023trivial}) with potential disorder (and quantum dots at ends in order to produce ZBCPs~\cite{liu2017andreev}). 
We note that the artificial features of the periodic pristine system in~\cite{hess2023trivial} disappear under realistic potential disorder. 
{The periodic model of Ref.~\onlinecite{hess2023trivial} with zero spin-orbit coupling and without realistic disorder is unjustified as a realistic approximation, and therefore, it cannot be used to interpret experimental results. We conclude by pointing out that the realistic disorder situation and its effects on Majorana nanowire experiments have already been extensively discussed in the literature~\cite{pan2021threeterminal,pan2020physical,dassarma2021disorderinduced,ahn2021estimating,liu2017andreev}, and Ref.~\onlinecite{hess2023trivial} does not add any substantive physics to the elucidation of disorder effects in Majorana nanowires through its artificial periodic Andreev band model.}
\bibliography{Paper_Comment}

\end{document}